\documentclass[twocolumn]{article}
\usepackage{graphicx}

\begin{document}
\title{Landau Level Crossings and Extended-State Mapping in Magnetic
Two-dimensional Electron Gases}
\author{R.\ Knobel and  N.\ Samarth \\
Department of Physics, Pennsylvania State University, University Park PA\\
16802\\
J.\ G.\ E.\ Harris and D.\ D.\ Awschalom\\
Department of Physics, University of California, Santa Barbara CA 93106}
\maketitle
\vspace{0.1cm}
\begin{abstract}
We present longitudinal and Hall magneto-resistance measurements of a
``magnetic'' two-dimensional electron gas (2DEG) formed in modul\-ation-doped
Zn$_{1-x-y}$Cd$_{x}$Mn$_{y}$Se quan\-tum wells. The electron spin splitting is
temperature and magnetic field dependent, resulting in striking features as
Landau levels of opposite spin cross near the Fermi level. Magnetization
measurements on the same sample probe the total density of states and Fermi
energy, allowing us to fit the transport data using a model involving
extended states centered at each Landau level and two-channel conduction for
spin-up and spin-down electrons. A mapping of the extended states over the
whole quantum Hall effect regime shows no floating of extended states as
Landau levels cross near the Fermi level.
\end{abstract}

\vspace{0.5cm}

When a perpendicular magnetic field is applied to a two-dimensional
electron gas (2DEG), the electron energy levels become quantized into highly
degenerate Landau levels (LLs) that are broadened by the presence of
disorder. Each broadened LL has a ``mobility edge'' separating localized
states in the tail regions from extended states in the central region of the
LL. The interplay between the extended and localized states in this simple
physical system has given rise to rich physics, ranging from the integer
quantum Hall effect (IQHE) \cite{prange90} to the magnetic field induced
metal-insulator transition.\cite{MIT}\cite{yuliarapcom} The location
of the mobility edges within the density of states (DOS) for each LL is
still a subject of research and speculation.\cite{furlan98} Particularly
interesting aspects of 2DEG physics are anticipated when circumstances allow
for the overlap of different LLs and their possible mixing. For instance,
calculations predict that LL mixing at vanishing magnetic fields may be
responsible for a ``floating'' of extended states to higher energies,\cite
{haldane97} hence offering a possible resolution to the magnetic-field
induced insulator-to-metal transition in 2DEGs.\cite{floating}\cite
{floating2} However, clean experimental studies of the consequences of LL
mixing at small magnetic fields are problematic since the existence of LLs
requires finite magnetic fields. This has motivated investigations of 2DEGs
in which LL crossings are purposely engineered away from zero magnetic
fields in the regime of quantum transport.\cite{hwj99}\cite{zeitler01}

Here, we use a ``magnetic'' 2DEG system (modulation-doped ${\rm {\
Zn_{1-x-y}Cd_{x}Mn_{y}Se}}$ quantum wells) to create finite field crossings
between opposite {\it spin} states of LLs with different orbital quantum
number. We observe striking anomalies in the magnetic field- and
temperature-dependence of the sheet and Hall resistance ($\rho_{xx}$ and $
\rho_{xy},$ respectively), including a negative differential Hall
resistance. Surprisingly, the data can be explained in detail using a simple
model of the extended states at the center of each disorder broadened LL,
without resorting to explicit LL mixing of the opposite spin states.
Further, by combining magnetization, optical and transport measurements on
samples from the same wafer, we map out the extended states in the LLs of
the 2DEG, providing for the first time a means to determine the form of the
extended states well away from a quantum Hall plateau. Our measurements
indicate that, at least in the present 2D system, LL crossings at finite
fields do not result in a floating of extended states.

Samples used in this study are grown by molecular beam epitaxy (MBE) and
consist of a modulation-doped ${\rm {Zn_{1-x-y}Cd_{x}Mn_{y}Se}}$ quantum
well located between ZnSe barriers. In these ``magnetic'' 2DEGs, the $s-d$
exchange interaction between confined electrons and local moments (Mn$^{2+}$
) results in a temperature- and magnetic field-dependent amplification of
the ``bare'' Zeeman energy. \cite{davidnitin} The ensuing LL fan diagram has
an unusual structure in which the dominance of the spin-splitting ($\Delta
E_{S}$) over the Fermi energy ($E_{F}$) and the cyclotron splitting can
result in a highly spin-polarized 2DEG even at low magnetic fields. Earlier
studies of low density samples (where $\Delta E_{S}\geq E_F$) have
shown how these can serve as model examples of a ``spinless'' 2D fermion gas
because all the LLs below $E_F$ are in the same spin state. \cite
{yuliarapcom} \cite{yuliaprl} \cite{ep2ds} The samples studied
here belong to a qualitatively different regime in which $\Delta E_{S}\simeq
E_F$, resulting in discrete points where Landau levels of opposite spin
cross near $E_F$.

In order to allow for complementary magnetometry measurements on the same
sample (reported in more depth else\-where \cite{jack}), an MBE-grown (100)
GaAs/\-GaAlAs heterostructure is used as a substrate. This paper focuses on a
sample in which the active 2DEG layer is deposited on a 1 $\mu $m thick ZnSe
buffer layer and consists of a modulation-doped 10.5 nm ($\sim 35$
monolayers) quantum well region in which $1/16$ monolayer of MnSe is
inserted every 7 monolayers of ${\rm {Zn_{1-x}Cd_{x}Se}}$ (giving an average
Mn composition of about 0.9 percent). The doping is provided by 20 nm of
n-doped ZnSe:Cl separated by 12 nm of undoped ZnSe on either side of the
quantum well. We note that the data shown here is qualitatively similar to
that seen in other samples with similar composition.

Magneto-resistance measurements are carried out using standard dc techniques
(excitation current $<30$ nA) on mesa-etched Hall bars ($1600\times 400\mu $
m) at low temperature (330 mK $\rightarrow $ 7 K) in a pumped helium-3
cryostat in magnetic fields up to 8T. Electrical contacts to the 2DEG are
made by first removing the oxide from the surface with an etch of ${\rm {\
Na_{2}HSO_{3}}}$ and using diffused indium contacts annealed in a forming
gas atmosphere for 15 minutes. Electron density in the 2DEG is varied
electrostatically with an evaporated gold gate, separated by $\sim 1\mu $m
of spun-on insulator (benzocyclobutene). Additional measurements carried out
at lower temperatures in a dilution fridge confirm that the data does not
change in any qualitative manner at temperatures below 300 mK. Magnetization
is measured over a comparable range of temperature and magnetic field using
cantilevers fabricated out of the same wafer.\cite{jack} Finally,
magneto-photoluminescence measurements are used to deduce the spin splitting
of conduction band states.

Figures 1 (a) and (b) show the magnetic field- and temperature-dependence of
$\rho _{xx}$ and $\rho _{xy}$ at zero gate voltage. We see a strong positive
magneto-resistance at low fields which is larger for low temperature and low
carrier concentration, as is generally seen in diluted magnetic
semiconductors.\cite{yuliaprl}\cite{dietl} SdH oscillations are
visible above $\sim 0.7$ T in (a), while $\rho _{xy}$ (b) shows oscillations
at moderate fields before showing properly quantized Hall plateaus for the
lowest temperature (0.35 K - solid line). Deriving the electron
concentration $n$ from the slope of $\rho _{xy}$ at low field ($n\approx
2.8\times 10^{11}{\rm {cm}^{-2}}$), we assign filling factors $\nu =\frac{hn
}{eB_{\nu }}$ to the minima of the $\rho _{xx}$ oscillations occurring at a
field $B_{\nu }$, and find that $\nu $ increments by approximately 1 between
each oscillation, indicating that the spin degeneracy of the Landau levels
has been lifted. However the SdH oscillations are not periodic in $1/B$; in
other words assigning integral values to $\nu $ makes the product $\nu \cdot
B_{\nu }$ vary with magnetic field. Furthermore, Fig. 1(a) also shows that
the SdH minima shift to lower magnetic fields with increasing temperature
(bottom to top), as highlighted with arrows in the figure. We emphasize that
the magnetic field dependence of $\rho _{xy}$ is linear before the onset of
the quantum Hall effect, indicating that the carrier density does not change
with magnetic field.\cite{comment1} We do note that $n$ determined from the
slope of the low-field $\rho _{xy}$ is generally less than that used in the
model developed later, and increases slightly as temperature increases. We
attribute this discrepancy in part to the anomalous Hall effect ($\rho
_{xy}=R_{H}B+R_{Anom}M$, where $M$ is the magnetization) and also to the
presence of some parallel conduction above 2.5 K; neither of these affect
the positions of the SdH oscillations.

In Fig.\ 1(a), we have circled a small oscillation around 3.2 T ($\nu =4$),
preceded by a large peak and accompanied by a distortion of the plateau in $
\rho _{xy}$ at 3 T. The minimum in $\rho _{xx}$ at 3.2 T is seen up to
higher temperatures (4.2 K), while the large minimum at 3.7 T disappears as
temperature is increased. A reduction in carrier density (and hence $E_F$)
using electrostatic gating rapidly destroys the small oscillation (Fig. 2),
and makes the SdH oscillations more regular while also dramatically
increasing the low field positive and high field negative
magneto-resistance. The aperiodicity in inverse magnetic field and shifts
with temperature observed in the SdH oscillations are also seen in de Haas
-- van Alphen (dHvA) oscillations measured using cantilever magnetometry (as
shown in Fig. 3(a)).

In order to understand these phenomena, we develop a model based on the LL
energy spectrum. In a magnetic 2DEG, the energy of an electron in a LL with
spin up (+) or down (-) is given by:
\begin{equation}
E_{\ell \pm }=\left( \ell +\frac{1}{2}\right) \hbar \omega _{c}\pm \frac{1}{2
} \Delta E_{S},
\end{equation}
where $\ell =0,1,2,\ldots$ is the level index, $\omega _{c}$ is the
cyclotron frequency, and $\Delta E_{S}$ is the spin splitting which follows
the empirical form\cite{gaj79}:
\begin{equation}  \label{equ:split}
\Delta E_{S}=\Delta E_{S_{MAX}}B_{5/2}\!\!\left[ \frac{5\mu _{B}B}{
k_{B}(T+T_{0})}\right] .  \label{2}
\end{equation}
Here, $\Delta E_{S_{MAX}}$ is the saturation value of the spin splitting, $
B_{5/2}$ is the spin-5/2 Brillouin function and $T+T_{0}$ is an effective
temperature. The Fermi energy is set by the constant number of electrons in
the quantum well $n$ and is calculated by integrating the DOS
$g(\varepsilon,B,T)$ with the Fermi-Dirac distribution function
$f(\varepsilon,T)$, and is calculated by numerically integrating
\begin{equation}
n=\int_{-\infty }^{\infty }g(\varepsilon ,B,T) f(\varepsilon,T)
d\varepsilon.
\label{equ:fermi}
\end{equation}

We model the DOS for electrons in the 2DEG using the energy levels of Eq.1
with a broadening due to disorder of width $\Gamma (B)$. This model is used
to fit the magnetization of the 2DEG, which is proportional to the
derivative of the free energy with respect to $B$, and thus is sensitive to
the DOS. The best fit parameters at 320 mK (Fig. 3(a)) of $\Delta E_{S_{{\rm
MAX}}}=7.7$ meV, $T_{0}=1.55$ K and $n=2.94\times 10^{11}$ cm$^{-2}$ are in
good agreement with magneto-photoluminescence measurements and low field
Hall measurements. The broadening is assumed to be a gaussian, and fit with
a width of $\Gamma =0.36B^{1/2}$ meV. An example of the resulting DOS (at T
= 0.32 K) is shown as a gray-scale plot in Fig. 3(b). Figures 3(c) and (d)
show how the LL energy diagram changes with temperature. Remarkably, as
shown for a single temperature in Fig. 3(a) and shown in extensive detail
elsewhere,\cite{jack} this simple form reproduces the features of the field-
and temperature-dependence of the magnetization without any additional
parameters.

The extension of this calculation to simulate the magneto-resistance
requires choosing a specific form for the density of extended states $
g_{ext}^{\ell s}(\varepsilon ,B,T)$. We model this density by assuming a LL
of index $\ell $ and spin $s$ has, at its center, a region of extended
states given by a gaussian of width $\Gamma _{ext}^{s}$ (independent of $B$
). This form phenomenologically accounts for a finite energy spread of the
extended states\cite{dassarma93} and variable-range hopping conductivity
through localized states. \cite{furlan98} The resistivity of each electron
spin channel is then calculated independently and is given by

\begin{eqnarray}
\rho _{xx}^{s }(B,T) &=&\rho _{0}\int_{-\infty }^{\infty
}\sum_{\ell}g_{ext}^{s \ell }(\varepsilon ,B,T)\frac{\partial f}{\partial
\varepsilon }(\varepsilon,T)d\varepsilon \nonumber \\
 & & \label{equ:rhoxx} \\
\rho _{xy}^{s }(B,T) &=&\frac{h}{\nu e^{2}}\nonumber \\
 &=&\frac{B}{\displaystyle
e\int_{-\infty }^{\infty }\sum_{\ell }g_{ext}^{s \ell
}(\varepsilon,B,T)f(\varepsilon ,T)d\varepsilon }  \label{equ:rhoxy}
\end{eqnarray}

where $f$ is the Fermi function and the total DOS and $E_F$ are determined
from the magnetization measurements. The sheet resistance $\rho _{xx}$
includes a prefactor $\rho _{0}$ which is assumed independent of the LL
index $\ell $ and the spin $s$. This model makes no attempt to account
microscopically for the strong positive (at low field) and negative (at
higher field) background magneto-resistance seen in all II-VI magnetic
semiconductors and attributed to bound magnetic polarons \cite{petukhov00}
and exchange-enhanced electron-electron interaction effects.\cite{yuliaprl}
The Hall resistance $\rho _{xy}$ is essentially calculated by counting the
number of LLs below $E_F$ and multiplying by the quantum of conductance.
The resistance $\rho $ is determined by adding the conductances $\sigma $
through each spin channel $s$ obtained by inverting the resistance tensor.
By using this form for the resistance, we implicitly assume that LL mixing
and floating of the extended states are unimportant in this regime, which
will be justified by the close agreement of the data to the model. Figures
1(c) and (d) show the simulated $\rho _{xx}$and $\rho _{xy}$at various
temperatures using $\Delta E_{S_{MAX}}$= 7.2 meV, $T_{0}$= 1.55 K, $
n=2.9\times 10^{11}$cm$^{-2}$, and $\Gamma $= 0.36 meV T${^{-1/2}}$ -- all
in good agreement with the parameters used for the magnetization. These
parameters model the data well over the whole temperature range, without
changing any other parameter. The model also provides a physical basis for
understanding the various anomalies observed in the transport data. First,
the unusual aperiodicity of the SdH oscillations arises simply from the
variable overlap of the extended states as LLs for spin up and down
electrons cross; the temperature-dependent shift of the SdH extrema can be
traced to the temperature variation of the spin splitting (Eq. 2) which
changes the LL spectrum. The negative differential Hall resistance seen at
low magnetic fields is a consequence of the addition of conductivities of
the two spin channels for conduction. The oscillation at 3.2 T is due to the
$|0,\uparrow >$ LL crossing $E_F$ near the field where $|3,\downarrow >$
LL also crosses.\cite{jaroszynski}  As temperature is increased, the spin up
level crosses at a higher field due to the decreased spin splitting,
allowing the oscillation to be more easily resolved (see Figs. 3(c) and(d)).
This oscillation is much smaller than those due to spin down LLs because
there are fewer spin up LL
s below $E_F$, causing $\rho _{xy}^{\,\uparrow }$
to be large (Eq. 5). Hence, the spin up LLs contribute less to the overall
conductivity than spin down LLs. This effect also explains why the spin up
LLs crossing $E_F$ at lower field produce only a slight modulation of the
SdH oscillations. The width of the extended states is best fit using widths
of $\Gamma _{ext}^{\,\downarrow }\simeq 0.25$meV for spin down electrons and
$\Gamma _{ext}^{\,\uparrow }\simeq 0.05$meV for spin up electrons. This
difference suggests that the minority spin electrons are more localized than
the majority spin electrons and this may influence the low-field positive
and high field negative magneto-resistance. Decreasing the sheet density $n$
using the electrostatic gate lowers $E_{F}$away from the region of
the energy diagram where LL crossings occur. Hence, the anomaly at $\sim $3
T disappears and the overlap between LLs is decreased overall, leading to
more regular SdH oscillations.

The complementary measurements of magnetization and transport in this study
provide a powerful means to determine the shape of the distribution of the
extended states. In the preceding discussion, we assumed a specific analytic
form $g_{ext}$ for the extended states; however, we can use the resistivity
to approximately map out the conducting states directly. In the regime where
a spin down LL $|\ell ,\downarrow >$ crosses $E_{F}$, the total resistance
approximates the spin down channel resistance ($\rho _{xx}(B)\simeq \rho
_{xx}^{\,\downarrow }(B)$) and is proportional to the density of conducting
states convolved with the derivative of the Fermi function (Eq.4). Using the
knowledge of $E_{F}$ and the LL spectrum obtained from the magnetization, we
transform $\rho _{xx}(B)$ into $\rho _{xx}(E)$, where $E=E_{F}-E_{|\ell
,\downarrow >}$ is the energy separation of the center of the LL from $E_{F}$
. Thus, the combination of data from the two experiments allows a
spectroscopy of the conducting states as $E_{F}$ passes through a LL. Figure
4 shows the temperature dependence of $\rho _{xx}$ through the $
|2,\downarrow >$ LL (about 4.5 T). As the temperature is increased, the $
|0,\uparrow >$ LL drops below $E_{F}$ in this region, leading to a
displacement of the shoulder on this peak from low to high energy. The
vertical line in each plot indicates the position of the $|0,\uparrow >$
level, i.e. $E_{|0,\uparrow >}-E_{F}$at the point where $
E_{F}=E_{|2,\downarrow >}$, and clearly corresponds to the shoulder seen in
the $\rho _{xx}$ peak. As $|0,\uparrow >$ and $|2,\downarrow >$ become
degenerate near $E_{F}$, no shifting of the $\rho _{xx}$ peak is seen from
zero energy. Such a shift would be evidence for floating of the extended
states.\cite{floating2} The derivative of the Fermi function for each
temperature is plotted for comparison, showing the resolution with which
this floating of the extended states is excluded. This simple crossing of
two LLs of opposite spin, well separated from others, is quite different
than the case at low field where all LLs overlap.

Figure 4 also shows that -- at low temperatures -- a simple delta function
for the extended states is not sufficient to explain the data, nor is a
finite width to the extended states alone enough to reproduce the data.
Variable-range hopping, which gives an exponential form to the conductivity
away from the LL center,\cite{polyakov93} must be included to give the peaks
sufficient width with a shape which matches the data. This justifies the
phenomenological gaussian form for $g_{ext}^{\ell s}(\varepsilon ,B,T)$ used
earlier. In a sample whose LL structure does not change with temperature
(such as a GaAs/AlGaAs modulation doped quantum well), this method should
allow the determination of the full form of the conduction through variable
range hopping and extended state transport in the QHE regime.

In summary, measurements of the magneto-resistance and Hall effect in 2DEGs
where the spin splitting is comparable to the Fermi level show the effects
of accidental degeneracies as LLs cross near the Fermi level. The resulting
anomalous SdH oscillations are well described using a model involving
extended states centered at each Landau level and two-channel conduction for
spin-up and spin-down electrons. The complementary measurement of
magnetization and transport measurements on the same sample allow us to
probe the location of the extended states away from the SdH minima. The
crossing of the LLs at large field does not induce a shifting of the
positions of the extended states -- which one might expect for ``floating''
states.

We thank K. D. Maranowski and A. C. Gossard for growth of the GaAs/GaAlAs
heterostructure, J. Gupta and I. Malajovich for taking the optical data and
S. A. Crooker for help with numerical modeling. This work was supported by
NSF grants NSF DMR-0071888 and -0071977, ONR N00014-99-1-0077 and -0071, and
AFOSR \#F49620-99-1-0033.

\onecolumn

\begin{figure}[tbp]\centering
\includegraphics[width=4.9in]{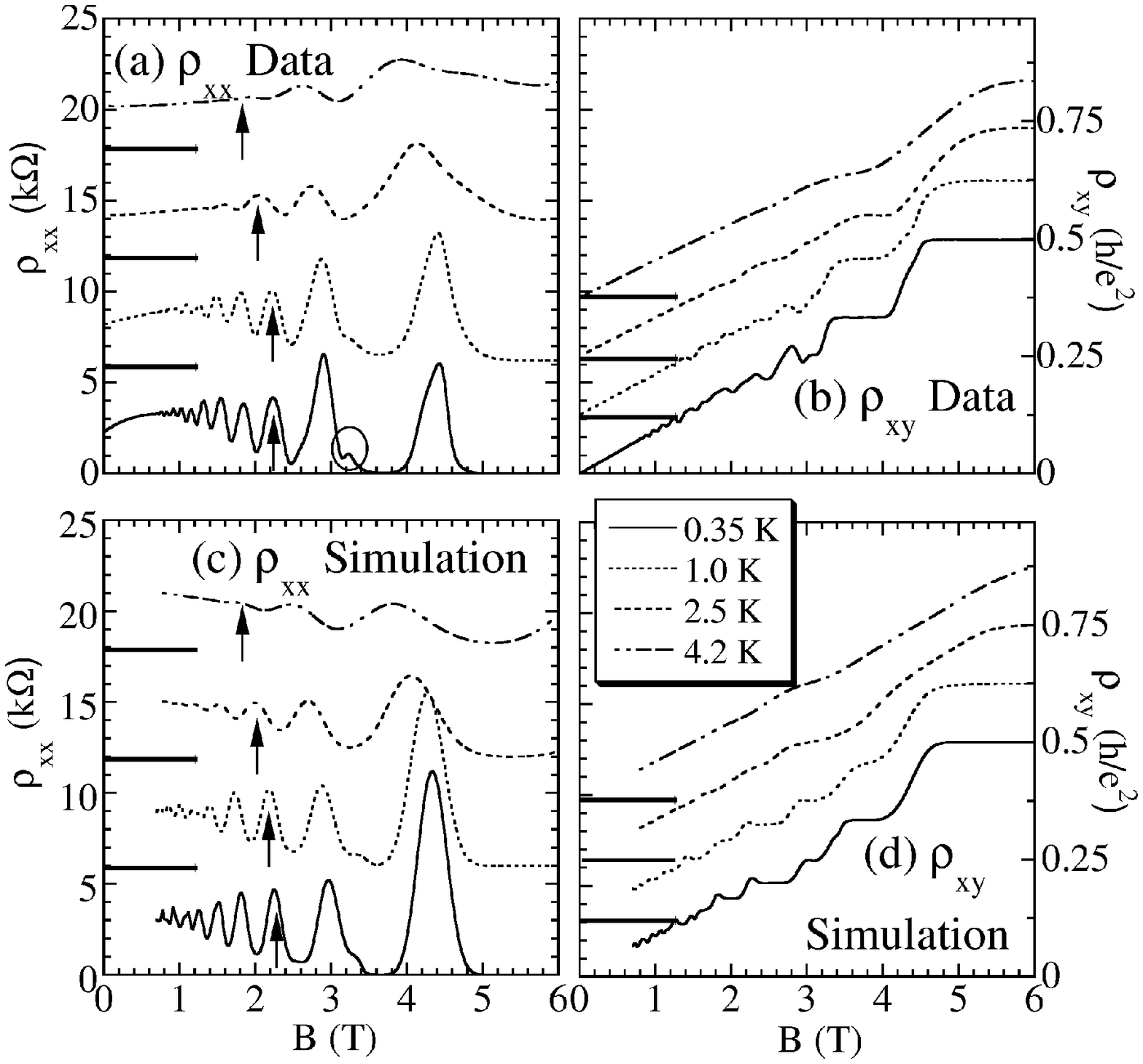}
\caption{Measured magnetic field- and temp\-era\-ture-depen\-dence of (a) $\protect
\rho _{xx}$ and (b) $\protect\rho _{xy}$ offset for clarity with the origin
for each trace indicated. The arrows in panel (a) emphasize how a particular
maximum in $\protect\rho _{xx}$ moves to lower field as temperature is
increased. The circle in panel (a) highlights the small dip corresponding to
filling factor $\protect\nu =4$ which becomes more pronounced at the expense
of $\protect\nu =3$ with increasing temperature. Panels (c) and (d) show
simulations of the magnetic field- and temperature-dependence of $\protect
\rho _{xx}$ and $\protect\rho _{xy}$, respectively, using the model
described in the text.}
\label{fig:temp}
\end{figure}

\begin{figure}[tbp]
\centering
\includegraphics[width=5.5in]{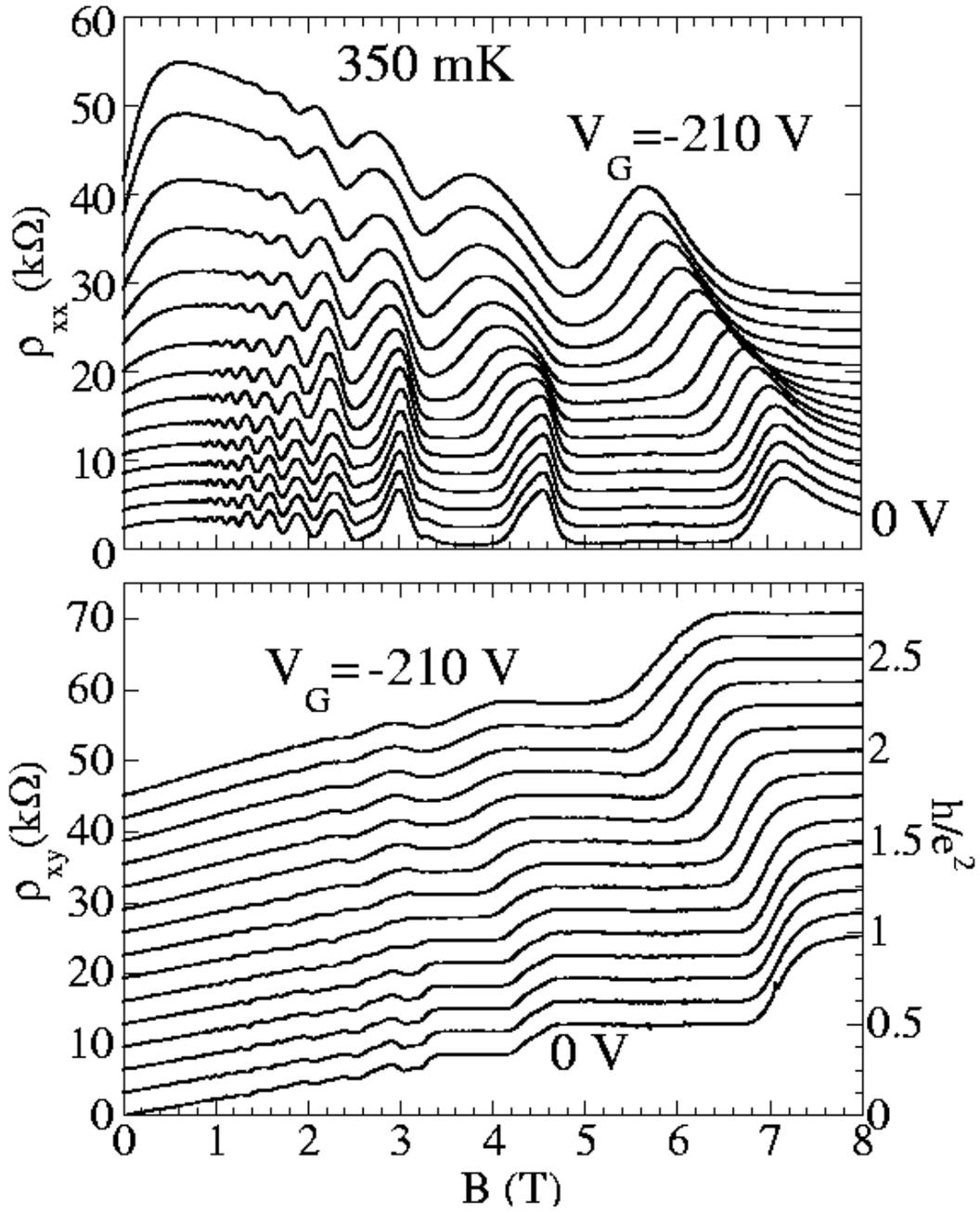}
\caption{Gate voltage dependence of (a) $\protect\rho_{xx}$ and (b)$
\protect\rho_{xy}$ at T = 350 mK offset for clarity. The gate voltage
increases from 0 to -210 V, decreasing the sheet density from 2.92 $\times
10^{11}$ cm$^{-2}$ to $\sim 2 \times 10^{11} {\rm cm^{-2}}$.}
\label{fig:gating}
\end{figure}

\begin{figure}[tbp]
\centering
\includegraphics[width=4.8in]{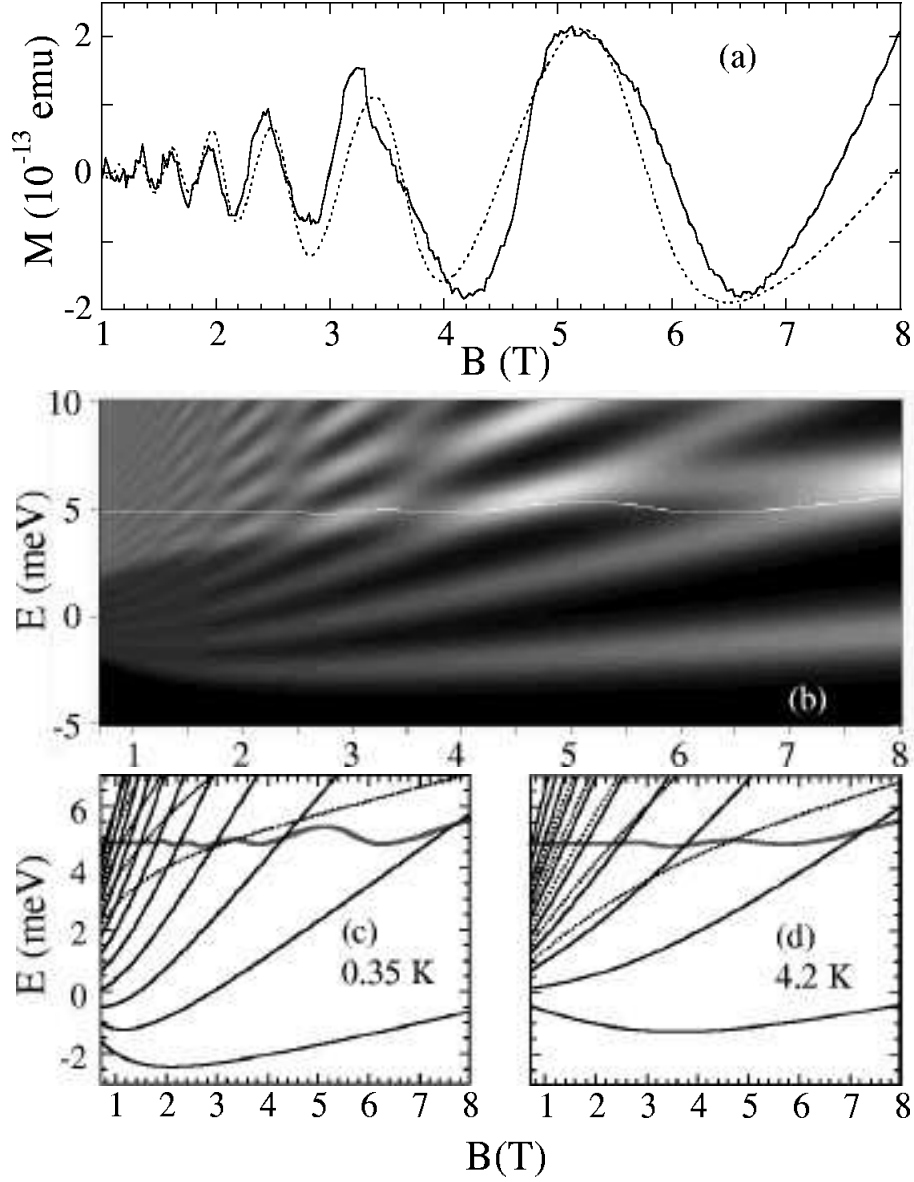}
\caption{(a) Measured magnetization of the 2DEG (solid line) as a function
of magnetic field at T=320 mK. The paramagnetic Mn$^{2+}$ background
magnetization has been subtracted. The dashed line is a fit to the data as
described in the text with the parameters $\Delta E_{S_{MAX}}$ = 7.7 meV, $%
T_0$ = 1.55 K, $n=2.94 \times 10^{11}$ cm$^{-2}$, and $\Gamma$ = 0.36 meV T$%
^{-1/2}$. (b) Gray scale plot of the total density of states as calculated
for the same model at T = 350 mK, with the Fermi energy $E_F$ indicated in
white. Panels (c) and (d) show the centers of the Landau levels and Fermi
energy for T = 350 mK and T = 4.2 K for this model.}
\end{figure}

\begin{figure}[tbp]\centering
\includegraphics[width=5.6in]{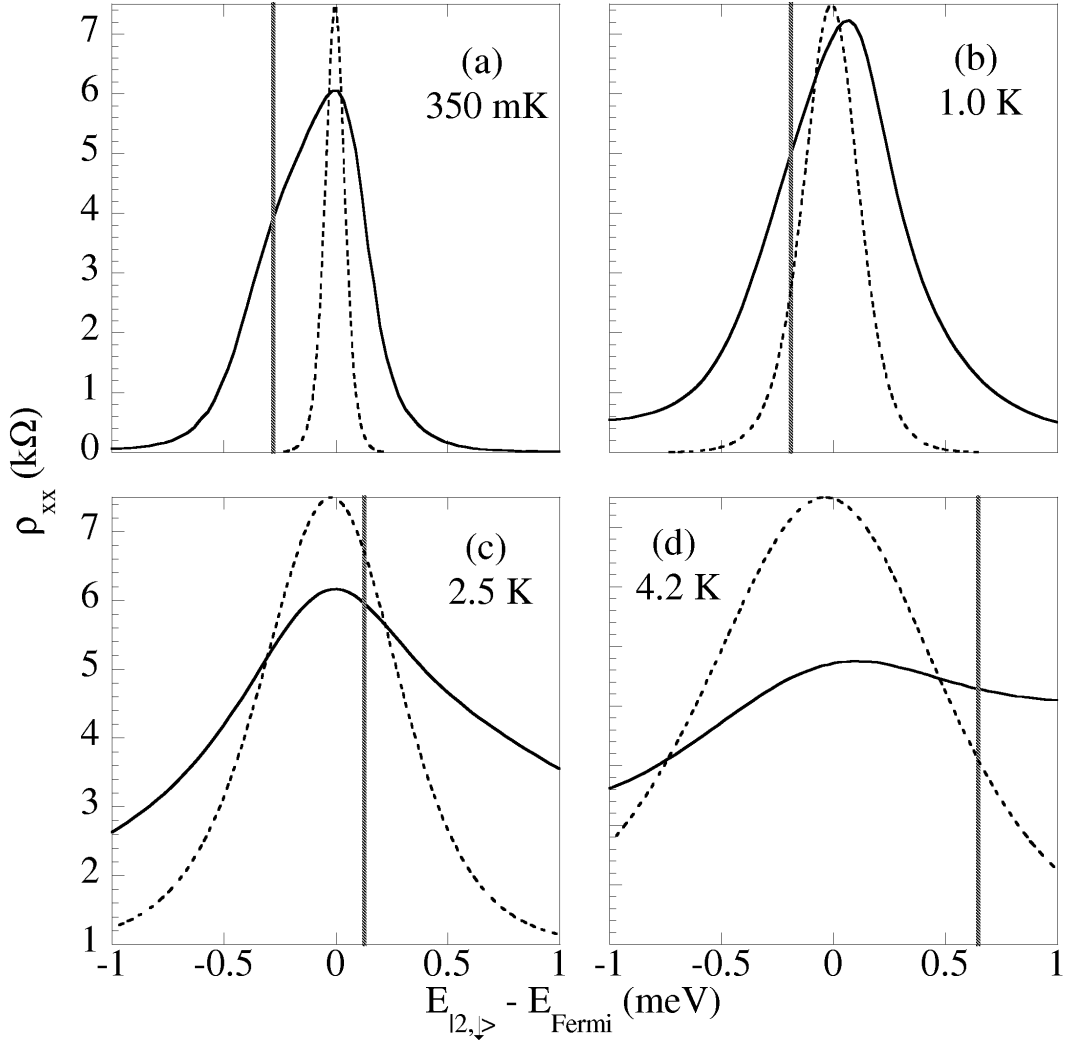}
\caption{Mapping of the extended states at four different temperatures. Each
panel shows the measured $\protect\rho _{xx}$\ as a function of $
(E_{|2,\downarrow \rangle }-E_{F})$ in meV, where $E_{|2,\downarrow \rangle
} $ and $E_{LL}$ are calculated for the parameters in Fig. 1 (c,d). The
dashed lines in each plot show the derivative of the Fermi function at each
temperature, indicating the broadening that would be expected for a delta
function density of extended states. The vertical line in each figure is the
position of the lowest spin up level $E_{|0,\uparrow \rangle }$ at the point
where $E_{|2,\downarrow \rangle }-E_{F}=0$.}
\label{fig:extended}
\end{figure}

\end{document}